\documentclass[12pt,preprint]{aastex}

\slugcomment{Scheduled to appear in the Feb. 2005 AJ}

\shorttitle{WHAM Point Sources}
\shortauthors{Reynolds et al}

\begin{document}

\title{Unresolved H$\alpha$ Enhancements at High Galactic Latitude in the
WHAM Sky Survey Maps}

\author{R. J. Reynolds,  V. Chaudhary, G. J. Madsen, \& L. M. Haffner}
\affil{Department of Astronomy, University of Wisconsin-Madison, Madison 
WI 53706}

%% \email{aastex-help@aas.org}

\begin{abstract}

We have identified 85 regions of enhanced H$\alpha$ emission at $|$b$| >
10\arcdeg$ subtending approximately 1$\arcdeg$ or less on the Wisconsin
H$\alpha$ Mapper (WHAM) sky survey.  These high latitude ``WHAM point
sources'' have H$\alpha$ fluxes of $10^{-11}$ to $10^{-9}$ erg cm$^{-2}$
s$^{-1}$, radial velocities within about 70 km s$^{-1}$ of the LSR, and
line widths that range from less than 20 km s$^{-1}$ to about 80 km
s$^{-1}$ (FWHM).  Twenty nine of these enhancements are not identified
with either cataloged nebulae or hot stars and appear to have kinematic
properties that differ from those observed for planetary nebulae.  
Another 14 enhancements are near hot evolved low mass stars that had no
previously reported detections of associated nebulosity.  The remainder of
the enhancements are cataloged planetary nebulae and small, high latitude
H~II regions surrounding massive O and early B stars.

\end{abstract}


\section{Introduction}

The Wisconsin H$\alpha$ Mapper (WHAM) has provided the first large-scale
survey of the distribution and kinematics of ionized interstellar
hydrogen, covering the sky north of declination $-30\arcdeg$ with and
angular resolution of about 1$\arcdeg$ and a velocity resolution of 12 km
s$^{-1}$ within approximately $\pm 100$ km s$^{-1}$ of the LSR (Haffner et
al 2003).  This survey shows interstellar H$\alpha$ emission filling the
sky, with loops, filaments, and other large emission enhancements
superposed on a more diffuse background.  However, in addition to these
large-scale features, the survey also reveals numerous small H$\alpha$
emission regions that have angular sizes comparable to or less than WHAM's
1$\arcdeg$ diameter beam.  In narrow ( $\approx 20$ km
s$^{-1}$) velocity interval maps, these WHAM point sources stand out
as intensity enhancements in a single beam (or two adjacent beams) within
a region of fainter diffuse emission.

Below we briefly describe our procedure for identifying and characterizing
these enhancements, and we list the resulting flux, radial velocity,
and line width of the H$\alpha$ emission, along with any previously
cataloged nebulosity or hot star that may be associated with the region.  
The nature of most of these emission regions is unknown.

\section{Identification of ``WHAM Point Sources''}

The enhancements were identified through a systematic search ``by eye''
through the entire data cube of the WHAM survey.  This consisted of
examining regions of the sky approximately 100 to 400 square degrees in
size within narrow (20 to 30 km s$^{-1}$) radial velocity intervals
centered between $-90$ km s$^{-1}$ and $+90$ km s$^{-1}$ (LSR).  To
minimize confusion with structure within bright, larger scale emission
features near the Galactic midplane, we confined the search to Galactic
latitudes $|$b$| > 10\arcdeg$. We also avoided the radial velocity
interval $-15$ to $+15$ km s$^{-1}$ in directions toward the
Orion-Eridanus bubble, where relatively bright high latitude H$\alpha$
emission features make the identification of ``point sources''
unreliable.  A less subjective identification program was also carried
out, which calculated for each of the approximately 37,000 survey spectra
the difference between the spectrum in a given direction and the average
spectrum of that direction's nearest neighbors.  Directions with an
enhancement were then selected based upon whether the difference spectrum
exhibited an emission feature that was significantly greater than the
scatter in the intensities of the nearest neighbors within the velocity
range of the feature.  This second method yielded a factor of ten more 
``point sources identifications''.  However, a cursory examination 
revealed that the vast majority of these were false positives associated 
with small angular scale fluctuations within the diffuse H$\alpha$ 
background.  We concluded that confidence in the identification of a true 
enhancement above the background required an examination ``by eye'' of a 
relatively large ($\sim 10\arcdeg \times 10\arcdeg$) region of the 
surrounding sky, not just the six nearest neighbors.  This allowed 
us to select only those enhancements that stood out most clearly 
against the background and was thus the more conservative approach.  The
survey was examined by two of us (VC and RJR) independently, and the good
agreement between the two results suggests that the enhancement
identifications are robust, with H$\alpha$ surface brightnesses measured
down to about 0.3 R (1 R = $10^6$/4$\pi$ photons cm$^{-2}$ s$^{-1}$
sr$^{-1}$), corresponding to an H$\alpha$ flux of about 1 $\times 10
^{-11}$ erg cm$^{-2}$ s$^{-1}$ for sources subtending 1$\arcdeg$.  While
this H$\alpha$ flux limit is not particularly low for a planetary nebula
search, the surface brightness limit, corresponding to an emission measure
of about 1 cm$^{-6}$ pc, is well below that of most planetary nebula
searches.  The sensitivity of WHAM is further enhanced over low spectral
resolution imaging in cases where the enhancement is Doppler shifted with
respect to the often higher surface brightness emission associated with
the ubiquitous warm ionized component of the interstellar medium (see
below).

The H$\alpha$ flux, radial velocity, and line width associated with each
enhancement was measured by subtracting from the spectrum toward the
enhancement the average spectrum toward the nearest neighbors, and then
fitting the resultant H$\alpha$ emission line with a Gaussian profile.  
Examples are presented in Figures 1 through 4, which show for four
relatively faint enhancements the velocity interval beam map of an area
surrounding the enhancement, the spectra in the source direction and its
nearest neighbors, and the difference spectrum with the best-fit Gaussian
and residuals. The intensity enhancements in these examples range from
1.3~R (Fig. 1) down to 0.5~R (Figs. 2 \& 3) and, depending upon the
brightness of the diffuse H$\alpha$ background, produce moderate (Fig. 2)
to small (Fig.  4) increases in the total H$\alpha$ intensity on the sky.  
The high spectral resolution of the WHAM survey has made possible the
detection of sources (e.g., Fig. 4) that would be masked by the H$\alpha$
background and its variation in maps of total H$\alpha$ intensity.  For
these examples, there are no associations with previously reported nebulae
(i.e., listed either in SIMBAD or in Cahn, Kaler, and Stanghellini 1992).  
In one of these examples (Fig. 2), the enhanced emission appears to be
associated with a hot evolved low mass star (DA white dwarf), while for
the other three no cataloged hot star is associated with the ionized
region (see \S3).

Seven of the enhancements occupy two adjacent pixels on the sky (e.g.,
Fig. 4), rather than being confined to a single WHAM beam. With one
exception (WPS 6), we have assumed that these are situations in which the
emission region is located near the edge of a beam, extending into the
second beam.  In these cases, we summed the two spectra, and the
coordinates of the enhancement refer to the mean position of the two
beams.  For the two-pixel source WPS 6 at $l = 33\fdg8$, b $= -22\fdg1$
and $l = 34\fdg1$, b $= -21\fdg2$, the results in Table 1 are listed
separately for each beam because there is a significant velocity shift
associated with the enhancement between the two directions (Fig. 5),
suggesting two independent sources (see \S4)

\section{Results}

Table 1 lists in order of increasing Galactic longitude the identified
``WHAM point sources'' (WPS) and the results of the Gaussian fits to the
difference spectra.  Following the WPS number are the Galactic coordinates
of the center of the WHAM survey beam (for single pixel sources) plus the
H$\alpha$ flux, radial velocity, and line width (FWHM) for each of the
enhancements.  The enhancements WPS 6-1 and WPS 6-2, while in adjacent
pixels, are treated as two separate H~II regions (see \S 4), while WPS 65
is a planetary nebula exhibiting two resolved H$\alpha$ velocity
components (Recillas-Cruz \& Pi\c{s}mi\c{s} 1981).  The H$\alpha$ flux is
based on a calibration using those enhancements identified as planetary
nebulae (see Table 2) and for which H$\beta$ fluxes and reddening
measurements have been published (Cahn et al 1992). The errors in the
fitted parameters are dominated by the uncertainty in the baseline of the
difference spectrum and/or by the scatter in the data points of the
spectum (e.g., see Figs.  1 - 4).  Errors due to baseline uncertainty were
estimated by fitting each difference spectrum multiple times with
different fixed baselines. Errors due to scatter in the spectral data
points were determined by the standard deviation calculation carried out
by the least-squares Gaussian fitting program.  The listed errors for each
parameter represent the largest uncertainties determined by these methods.

Table 2 lists for each enhancement the celestial coordinates (2000.0) of
the center of the beam (or the mean position for two-pixel sources), the
name of any cataloged nebulae near that direction, the name of a candidate
ionizing star (if there is no cataloged nebula), the spectral type of the 
star, and the off-set of the nebula or star from the center of the beam.  
The resources for the ionizing star and nebula searches were SIMBAD and
the planetary nebula catalog by Cahn et al (1992).  When no cataloged
nebula was listed within 40$\arcmin$ of the beam center, a search for an
ionizing star was carried out on SIMBAD to a radius of $60\arcmin$.

\section{Discussion}

Of the 85 H$\alpha$ enhancements identified, more than half (44) are not 
associated
with any previously cataloged nebula, and of these, fifteen are associated
with hot evolved low mass stars, including one DO and seven DA white
dwarfs, three SdO, and two SdB stars (Table 2). This is a potential source
of new information about the natures of these evolved stars and their
evolution (Tweedy \& Kwitter 1994).  For example, because of WHAM's large
beam, some of these enhancements could be associated with large planetary
nebulae in very late stages of their evolution having surface brightnesses
that are too faint to have been detected on earlier searches.  The
identification of the nebulosity with the star is most certain for those
stars within the WHAM beam (i.e., angular offsets $ < 30\arcmin$).  
However, because large, highly evolved planetary nebulae can be offset
significantly from their ionizing stars (Tweedy \& Napiwotzki 1994;
Borkowski, Sarazin, \& Soker 1990; Reynolds 1985), we have considered
stars located up to 1$\arcdeg$ from the directions listed in the Tables.  
We confirm the earlier detection of ionized gas associated with the DO
white dwarf PG 0108+101 (Reynolds 1987) and provide improved kinematic
information about that H~II region.

Twenty nine emission regions could not be associated with either a
cataloged nebula or hot star.  This could be the result of incompleteness
in the SIMBAD listings, or it could indicate another kind of nebulosity.  
These enhancements have a mean line width near 27 km s$^{-1}$,
significantly smaller than that (38 km s$^{-1}$) of the cataloged
planetary nebulae.  This is illustrated in Figure 6, which compares
histograms of line widths for three categories of WHAM point sources:
enhancements not associated with any cataloged nebula or evolved hot star,
enhancements near a hot low mass star, enhancements associated cataloged
planetary nebulae (the six regions associated with massive O and B stars
are excluded from these histograms).  We found no associations with
supernova remnants or Herbig-Hero objects, although WPS 11 and WPS 21 are
within $37\arcmin$ and $48\arcmin$ of the two high Galactic latitude
molecular clouds, MBM 50 and MBM 46, respectively, which could harbor star
formation activity.  However, the narrow line widths appear to rule out
such shock excited sources, as well as any association with emission line
stars, which exhibit line widths in excess of 60 -- 100 km s$^{-1}$ (e.g.,
Hartigan et al 1987; Hamann \& Persson 1992a,b).  It is tempting to
speculate that these emission regions are associated with the most evolved
planetary nebulae, those whose expansion has been halted by interactions
with the ambient interstellar medium (Tweedy \& Napiwotzki 1994; Reynolds
1985), or those whose gas has thinned to such an extent that it is the
ambient interstellar medium itself that has become the primary H~II region
(Borkowski, Sarazin, \& Soker 1990).  Followup, high angular resolution
imaging of these regions could help to discriminate between the
possibilities (Soker, Borkowski, \& Sarazin 1991).  Figures 7 and 8 show
corresponding histograms for radial velocity and H$\alpha$ flux.  No clear
differences between the three catagories of enhancements are apparent in
these distributions. In Figures, 6, 7, and 8, the V$_{LSR}$ and FWHM used
for WPS 65 are the flux weighted average radial velocity (i.e., +19 km
s$^{-1}$) and the separation (i.e., 37 km s$^{-1}$) of the two velocity
components, respectively.

Six of the enhancements appear to be small H~II regions associated with
massive late O and early B type stars.  The region near the B1~V star
HD~191639 (WPS 6) was detected in two pixels and exhibits a significant
(21 km s$^{-1}$) radial velocity difference between the two directions
(Fig. 5).  This suggests either peculiar, small-scale kinematic variations
within the region or the existence of two independent H~II regions closely
spaced on the sky.  In this latter case, because the B star has a radial
velocity of $-7 \pm 5$ km s$^{-1}$ (Wilson 1953), the emission region
produced by the B star would be more likely associated with the
enhancement (WPS 6-1) at $-10 \pm 2$ km s$^{-1}$ toward $l = 33\fdg8$, b
$= -22\fdg1$, while the emission (WPS 6-2) at $+11$ km s$^{-1}$ toward $l
= 34\fdg1$, b $= -21\fdg2$ would have no identified source of ionization
(for Figs. 6, 7, \& 8, we have assumed this latter case).  Enhancements
WPS 23, 60, 68, and 72 have late B stars (B6~V, B9, B9, and B8/B9II,
respectively) located $34\arcmin$ to $43\arcmin$ away.  Because the Lyman
continuum fluxes from such late type B stars are predicted to be orders of
magnitude weaker than the fluxes of the early B stars discussed above, we
have concluded that these associations are coincidences.

\section{Summary and Conclusions}

From the WHAM sky survey we have identified and measured the fluxes,
radial velocities, and line widths for 85 regions of H$\alpha$ enhancement
at Galactic latitudes $|$b$| > 10\arcdeg$ that appear to subtend
approximately one degree or less on the sky.  Most of these ionized
regions have not been previously reported as emission nebulae, and their
nature is unknown.  A next step is to carry out additional observations to
determine the morphology of these emission regions and their sources of
ionization.  This will include spectra of [O~III] $\lambda$5007, [N~II]
$\lambda$6584, and [S~II] $\lambda$6716 to explore the ionization and
excitation state of the gas, as well as observations using WHAM's
``imaging mode'' to obtain deep, very narrow band (30 km s$^{-1}$) images
of these enhancements at an angular resolution at about $3\arcmin$ within
a $1\arcdeg$ field of view (Reynolds et al 1998).

\acknowledgments

We thank an anonymous referee for helpful comments.  This work was funded
by the National Science Foundation through grants AST96-19424 and
AS02-04973.  The WHAM survey was funded by the National Science Foundation
through grants AST 91-22701 and AST 96-19424 with assistance from the
University of Wisconsin's Graduate School, Department of Astronomy, and
Department of Physics.  This research has made use of the SIMBAD database,
operated at CDS, Strasbourg, France.

{}

\clearpage

\begin{figure}
\epsscale{.8}
\plotone{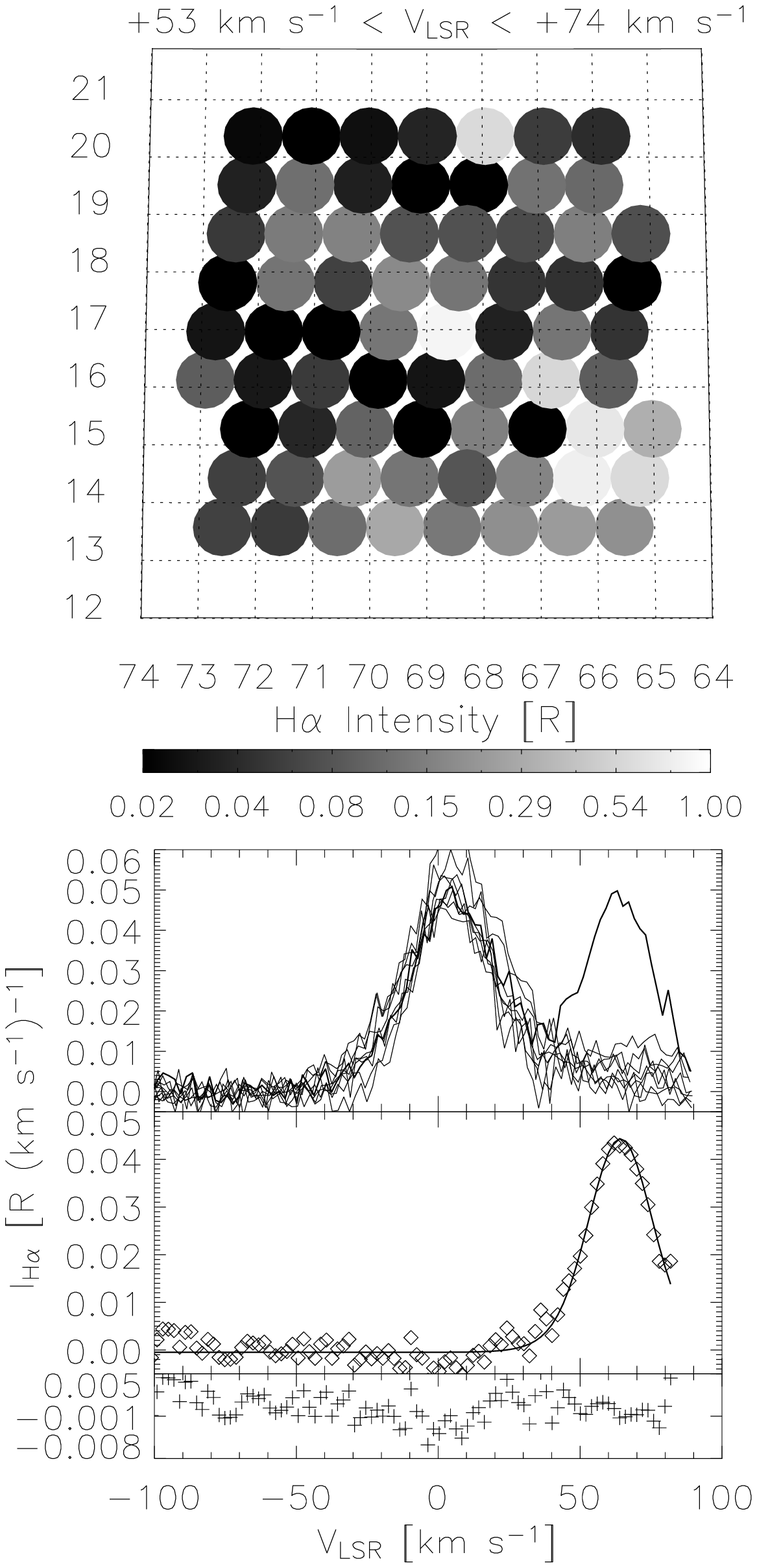}
\caption{Top: a radial velocity interval map of a small portion of the
WHAM H$\alpha$ sky survey showing the intensity enhancement toward WHAM
Point Source (WPS) 20, at l = 68$\fdg7$, b $= +17\fdg0$.  The circles
represent the WHAM beams and the grey scale denotes the H$\alpha$
intensity within each beam (in rayleighs).  The coordinates are degrees of
Galactic longitude (abscissa) and latitude (ordinate).  Bottom: H$\alpha$
spectra (solid lines) toward WPS 20 and its nearest neighbors; the
difference spectrum (diamonds) with its single-Gaussian best-fit and
residuals (crosses); see \S2. in text.  The H$\alpha$ line centered
near 0 km s$^{-1}$ in all of the spectra is emission associated with the
widespread warm ionized component of the interstellar medium (e.g., 
Haffner et al 2003).}
\label{fig1}

\end{figure}

\clearpage

\begin{figure}
\epsscale{1}
\plotone{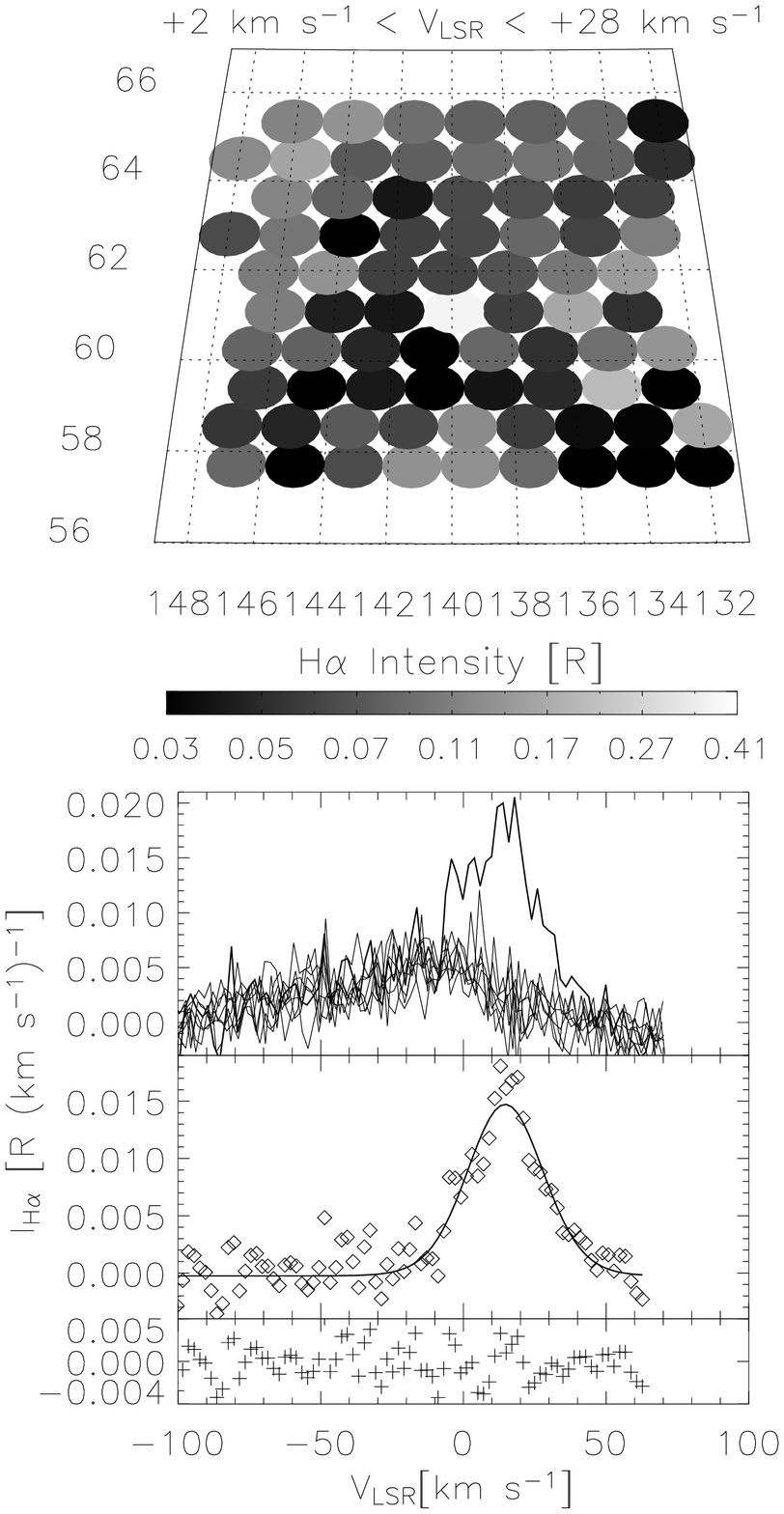}
\caption{Same as Fig 1. except for WPS 43.}
\label{fig2}
\end{figure}

\clearpage

\begin{figure}

\plotone{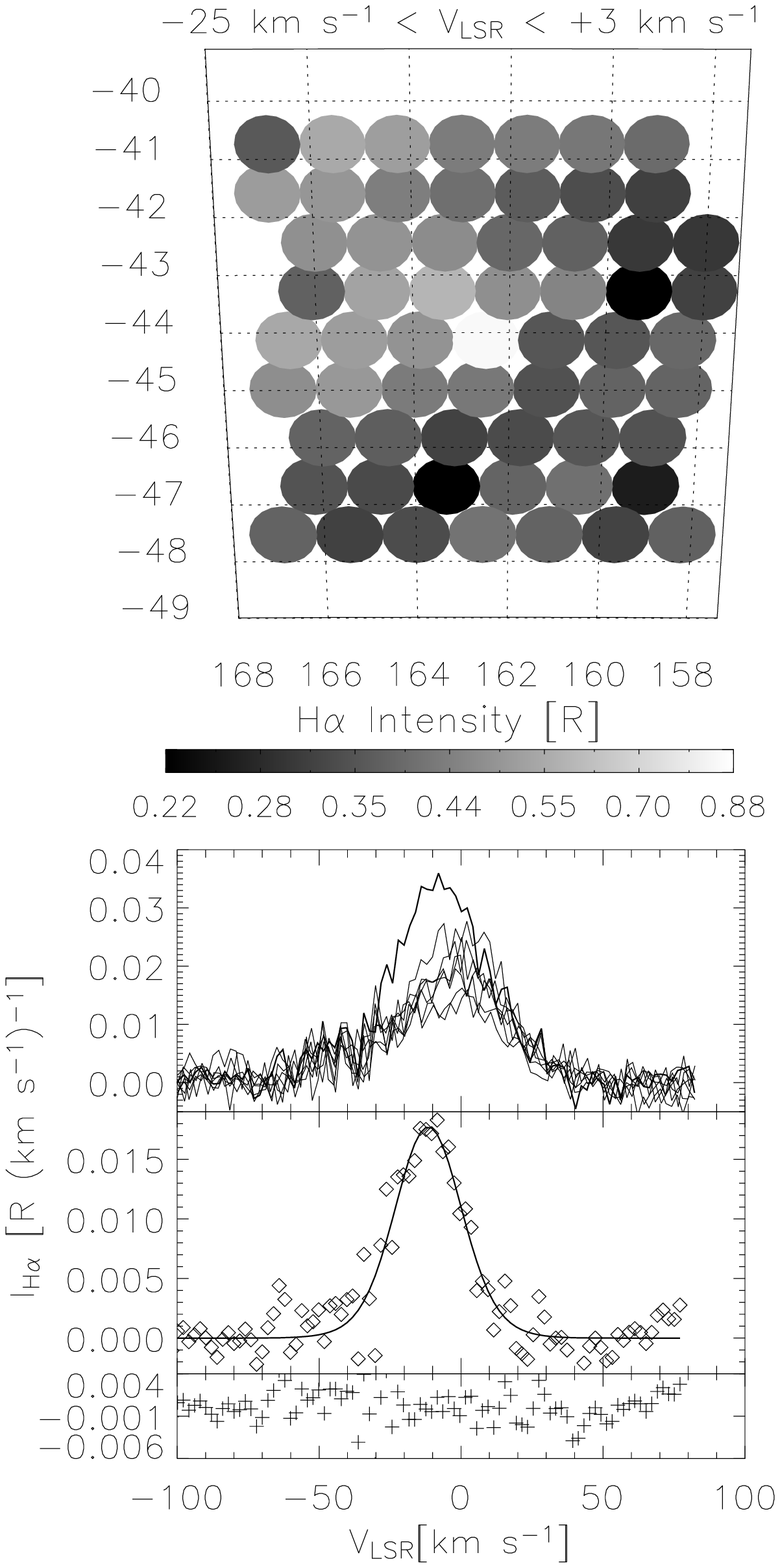}
\caption{Same as Fig 1. except for WPS 53.}
\label{fig3}
\end{figure}
\clearpage

\begin{figure}
\epsscale{.9}
\plotone{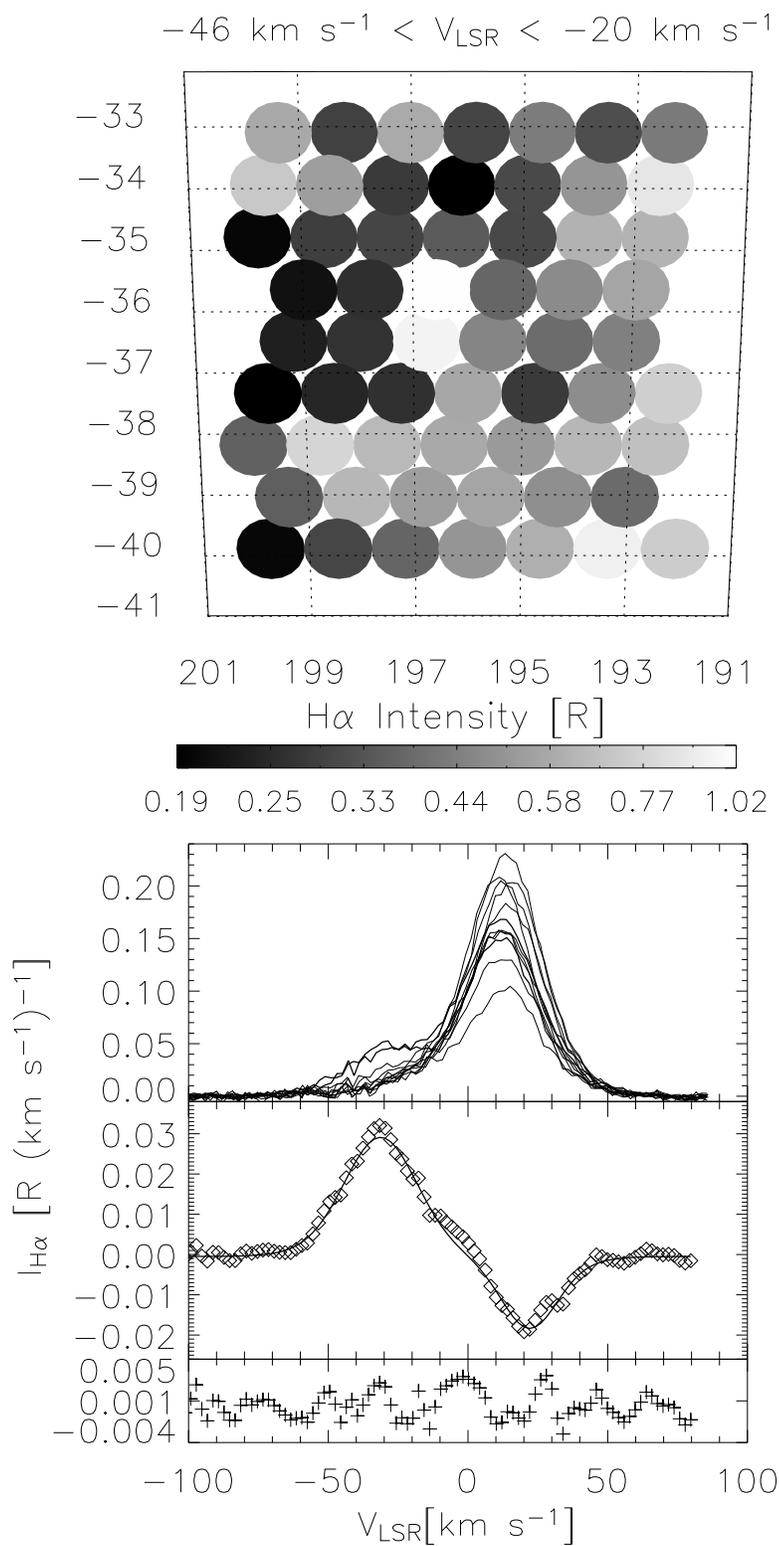}
\caption{Same as Fig 1. except for WPS 63.  Note that this is a 
two-pixel enhancement.  The parameters for this source listed in 
Table 1 refer to the positive-area Gaussian centered at about 
$-30$ km s$^{-1}$ in the difference spectrum.}
\label{fig4}
\end{figure}
\clearpage

\begin{figure}
\epsscale{.9}
\plotone{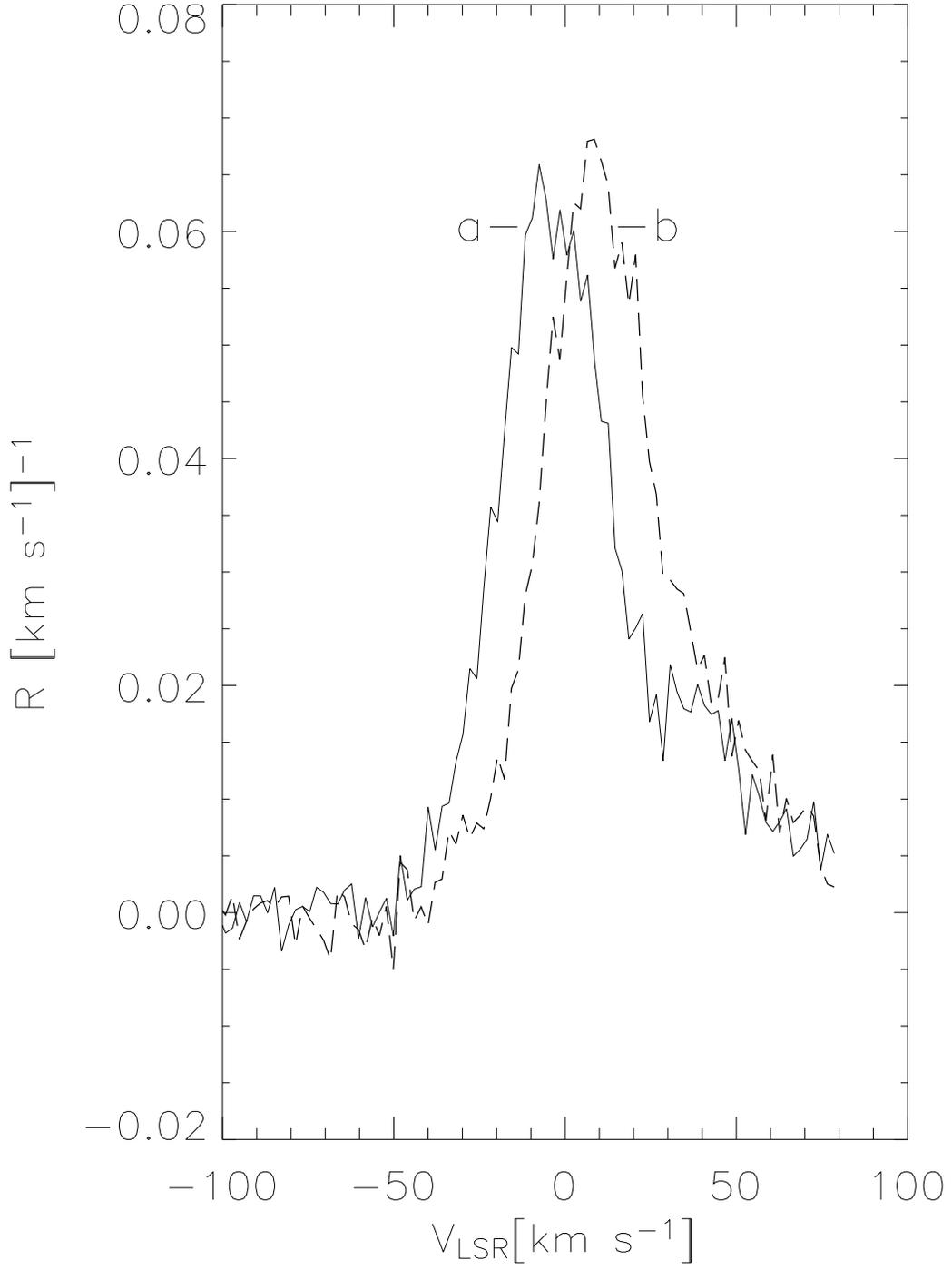}
\caption{H$\alpha$ difference spectra toward WPS 6-1 and WPS 6-2.  a: $l =
33\fdg8$, b $= 22\fdg1$; b: $l = 34\fdg1$, b $= -21\fdg2$.}
\label{fig5}
\end{figure}
\clearpage

\begin{figure}
\epsscale{.9}
\plotone{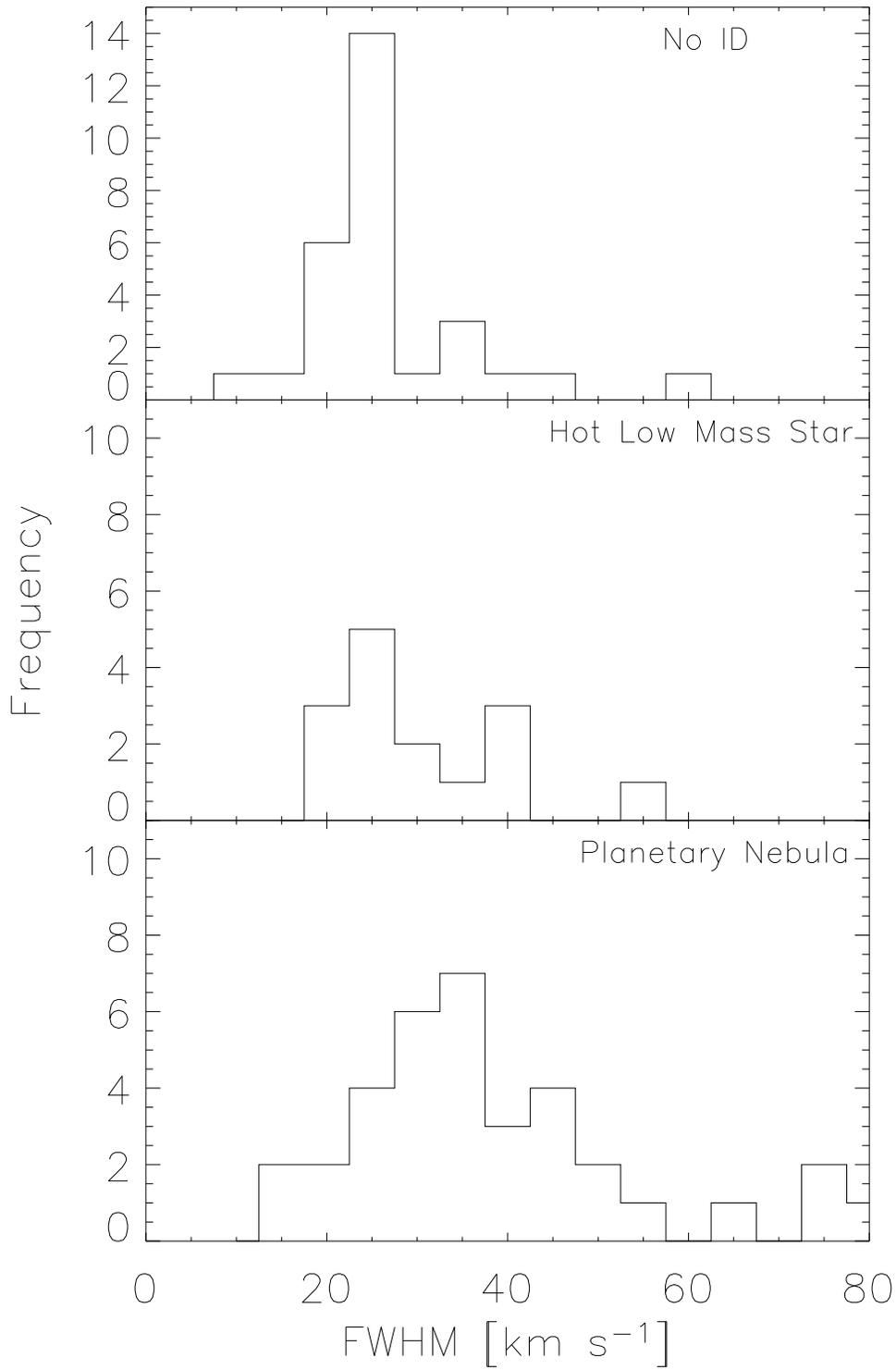}
\caption{Histograms of the number of enhancements versus FWHM of the
H$\alpha$ emission line.  Top: enhancements not associated with any
previously catalogued nebula nor any identified ionizing source. Middle:
enhancements not associated with any previously catalogued nebula but with
a hot evolved low mass star located within one degree of the center of the
WHAM beam.  Bottom: enhancements associated with catalogued planetary
nebulae.}
\label{fig6}
\end{figure}
\clearpage

\begin{figure}
\epsscale{1}
\plotone{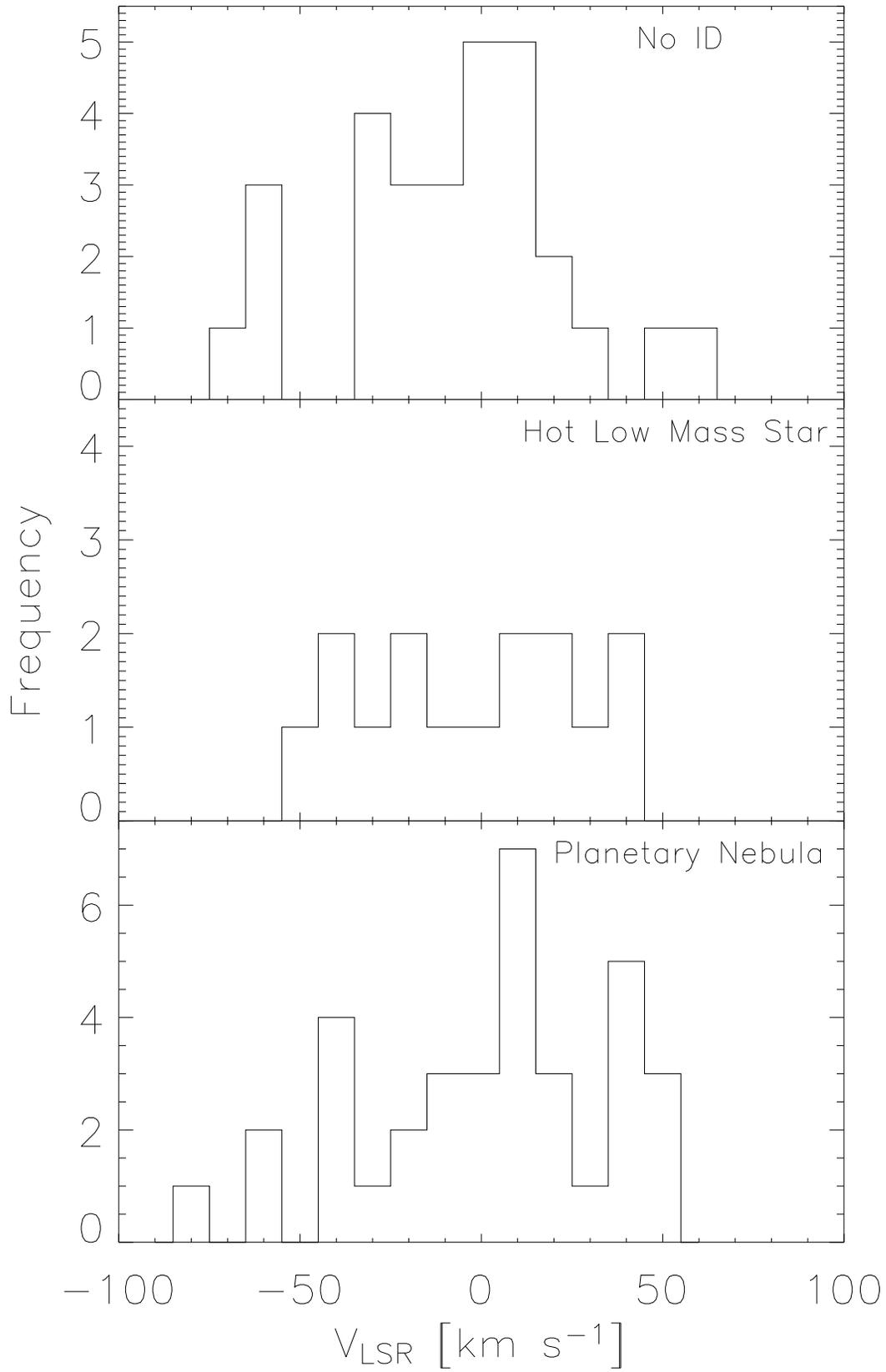}
\caption{Same as Fig. 6, except histograms of the radial velocity of the
H$\alpha$ emission.}
\label{fig7}
\end{figure}
\clearpage

\begin{figure}

\plotone{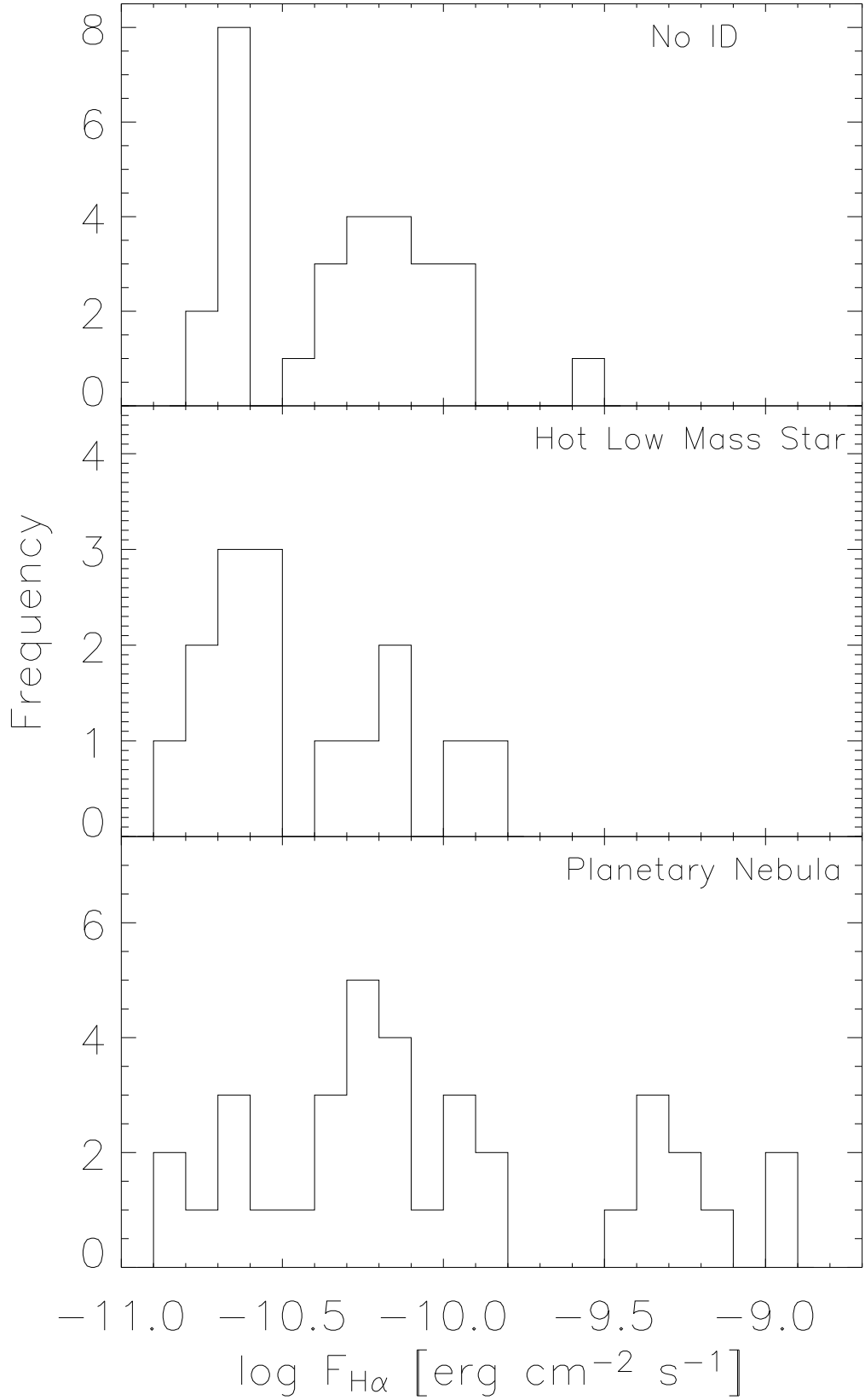}
\caption{Same as Fig. 6, except histograms of the flux of the H$\alpha$
emission.}
\label{fig8}
\end{figure}
\clearpage

\begin{deluxetable}{cccr@{$\pm$}lr@{$\pm$}lr@{$\pm$}l}

\tablecaption{Results for High Latitude Wham Point Sources}
\tabletypesize{\small}
\tablewidth{0pt}
\tablenum{1}
\tablehead{ \colhead{WPS} & \colhead{l}& \colhead{b} & \multicolumn{2}{c}{F$_{H\alpha}$} &  \multicolumn{2}{c}{V$_{LSR}$} &  \multicolumn{2}{c}{FWHM} \\
\colhead{} & \colhead{(deg)} & \colhead{(deg)} & \multicolumn{2}{c}{(10$^{-11})$ \tablenotemark{a}} & \multicolumn{2}{c}{(km s$^{-1}$)} & \multicolumn{2}{c}{(km s$^{-1}$)} } 
\startdata
1&  23.14& -17.83&  13.8&1.7&  +38&1&  39&3 \\
2&  23.76& -14.43&  8.8&0.8&   +47&1&  18&2 \\
3&  25.87& +40.74&  7.9&0.8&   +41&1&  29&3 \\
4&  26.20& +46.68&   2.9&1.3&   -4&6&  56&15 \\
5&  26.23& -17.83&  8.8&1.7&   +14&2&  43&4 \\
6-1\tablenotemark{b}&  33.84& -22.07&  5.0&0.8&   -10&2&  24&4 \\
6-2\tablenotemark{b}&  34.13& -21.22&  4.2&1.3&   +11&2&  25&5 \\
7&  35.05& +11.88&  56.8&1.3&  +11&1&  35&12 \\
8&  36.35& -56.86&  111.2&2.5& -24&1&  40&5 \\
9&  37.49& -34.80&  51.8&4.2&  -40&2&  45&1 \\
10& 38.06& +11.88&  5.4&0.8&   +39&2&  24&4 \\
11\tablenotemark{c}& 39.84& -35.23&  10.8&1.7&  +26&3&  25&5 \\
12\tablenotemark{c}& 43.01& +37.77&  40.1&1.7&  -20&1&  23&2 \\
13& 44.77& -46.68&  2.1&0.4&   -28&4&  34&6 \\
14& 45.76& +24.61&  2.9&0.8&     0&2&  22&6 \\
15& 54.08& -11.88&  10.9&2.1&  +48&3&  51&4 \\
16\tablenotemark{c}& 55.00& -14.86&  7.5&1.7&   +13&2&  21&3 \\
17& 57.03& +43.28&  8.4&4.2&   -69&4&  22&12 \\
18& 58.40& -11.03&  7.1&2.1&   -40&7&  39&15 \\
19& 65.56& +22.07&  2.9&0.8&   -37&3&  27&10 \\
20& 68.65& +16.97&  5.4&0.8&   +64&1&  21&3 \\
21& 69.31& -18.67&  6.3&0.8&   +52&2&  27&4 \\
22& 69.42& -30.55&  3.3&0.8&   +30&4&  36&14 \\
23& 71.38& +18.67&  9.2&0.8&   +17&1&  27&4 \\
24& 72.75& -16.97&  5.0&0.8&   +43&5&  23&7 \\
25& 74.85& -36.49&  4.6&0.8&   +10&3&  32&5 \\
26& 83.73& +72.99&  2.1&0.8&   -24&5&  37&13 \\
27& 83.88& +12.73&  47.2&2.9&  +13&1&  34&1 \\
28& 91.99& +19.52&  3.3&0.8&   +6&3&   27&6 \\
29& 96.39& +29.70&  78.6&2.9&  -55&1&  37&1 \\
30& 106.52& -17.82& 36.8&1.3&  -10&1&  49&2 \\
31& 109.82& +21.22& 7.5&0.4&   -22&1&  20&2 \\
32& 110.16& -11.88& 2.5&0.8&   +50&4&  31&9 \\
33& 118.60& +39.04&  2.5&0.8&   +15&3&  24&8 \\
34& 118.75& -74.69& 5.9&2.1&   34&10&  73&20 \\
35& 120.00& +18.67& 7.1&0.8&    +4&1&  25&1 \\
36& 123.41& +34.80& 5.4&0.8&   -35&2&  34&4 \\
37& 126.19& -46.68& 1.7&0.4&    +6&1&  20&3 \\
38& 130.75& -52.62& 2.1&0.8&    -14&3&  21&8 \\
39& 133.11& +12.73& 4.6&0.8&   -33&2&  27&5 \\
40& 133.22& -40.74& 1.3&0.4&   -36&4&  22&13 \\
41& 137.39& -24.61& 5.4&1.3&   -63&2&  26&3 \\
42& 139.24& +45.83& 2.1&0.8&   -57&7&  47&14 \\
43& 139.94& +61.11& 2.1&0.8&   +15&4&  26&11 \\
44& 142.11& -28.86& 5.4&1.3&   -26&4&  58&8 \\
45& 142.23& +15.28& 1.3&0.4&   -79&2&  13&4 \\
46\tablenotemark{c}& 143.86& +38.62& 5.0&0.8&   -52&2&  31&5 \\
47& 146.70& -23.76& 2.9&0.8&   +30&4&  39&7 \\
48& 149.29& +56.86& 5.9&1.7&   +15&5&  79&11 \\
49& 159.02& +17.82& 5.9&0.8&    +7&2&  37&5 \\
50& 159.02& +37.34& 2.1&0.8&    -9&5&  28&13 \\
51& 160.62& -23.76& 11.3&3.3&   +9&1&  23&2 \\
52& 161.20& -26.31& 7.5&1.3&    +8&1&  22&3 \\
53& 162.48& -44.13& 2.1&0.4&   -11&3&  28&7 \\
54& 162.55& +47.53& 1.7&0.8&   -19&5&  38&14 \\
55& 163.94& -16.97& 13.8&1.3&   -6&1&  17&1 \\
56& 164.68& +31.40& 2.1&8.4&   -60&5&  44&15 \\
57& 165.59& -15.28& 4.2&0.8&   +42&4&  42&10 \\
58& 166.28& +10.18& 10.0&0.8&  -39&1&  31&2 \\
59& 174.73& +15.28& 10.0&1.3&  -33&1&  27&3 \\
60& 181.26& +11.88& 7.1&1.3&    -2&1&  24&3 \\
61& 191.42& +40.74& 1.7&0.4&    +5&7&  32&19 \\
62& 193.44& -24.61& 1.5&2.9&   -11&1&  14&2 \\
63\tablenotemark{c}& 196.66& -36.07& 9.2&1.7&   -33&1&  26&2 \\
64& 199.45& -22.92& 4.2&0.4&   -63&2&  23&4 \\
65\tablenotemark{d}& 204.90& +14.43&  7.5&0.8&   +2&3&  22&10 \\
65\tablenotemark{d}& 204.90& +14.43&  6.3&1.3&  +39&2&  28&8 \\
66& 211.31& +11.88& 6.7&0.7&    +3&1&  20&2 \\
67& 215.00& -24.61& 100.7&2.1& +44&1&  28&2 \\
68& 216.18& -67.89& 2.1&0.8&   -12&5&  34&13 \\
69& 217.07& +28.86& 11.7&2.5&   -3&1&  20&2 \\
70& 218.64& +31.40& 4.6&1.3&   +22&3&  54&7 \\
71& 220.09& -54.31& 11.3&1.3&  +49&3&  73&5 \\
72& 223.71& -20.37& 29.3&1.7&   -7&1&  17&1 \\
73& 235.12& +47.53& 1.7&0.4&  +16&5&  35&12 \\
74& 235.17& +29.70& 5.4&0.8&    +6&2&  25&5 \\
75& 243.09& +36.49& 2.5&0.4&     0&3&  24&7 \\
76& 260.73& +32.25& 48.1&2.5&   +1&1&  43&1 \\
77& 264.29& +50.92& 1.7&0.4&    -3&5&  26&10 \\
78& 293.97& +43.28& 7.5&1.3&   +12&7&  67&15 \\
79& 303.20& +39.89& 15.1&0.4&  -12&1&  28&2 \\
80& 312.12& +67.05& 2.5&0.8&   -23&2&  12&7 \\
81& 315.00& +36.49& 2.5&0.8&   +39&5&  40&14 \\
82\tablenotemark{c}& 337.44& +47.53& 6.7&1.7&   -17&2&  26&5 \\
83& 347.06& +20.37& 219&29&     +1&1&  21&1 \\
84& 351.45& +16.97& 493&50&      0&1&  20&1 \\
\enddata
\tablenotetext{a}{erg cm$^{-2}$ s$^{-1}$.}
\tablenotetext{b}{Two-pixel point source where the radial velocity is 
different in each beam; see text.}
\tablenotetext{c}{Denotes two-pixel point source.}
\tablenotetext{d}{One-pixel point source with 2 emission peaks.} 
\end{deluxetable}

\clearpage

\begin{deluxetable}{ccccccc}
\rotate
\tablecaption{Wham Point Source Identifications}
\tabletypesize{\small}
\tablewidth{0pt}
\tablenum{2}
\tablehead{ \colhead{WPS} & \colhead{RA} & \colhead{Dec} & \colhead{Nebula} & \colhead{Star} & \colhead{Spectral Type} & \colhead{Beam Offset} \\
\colhead{ } & \colhead{ } & \colhead{ } & \colhead{ } & \colhead{ } & \colhead{ } & \colhead{(Arc Minutes)} }
\startdata
1&   19 39.3&  -16 29&     ...&               LSE 22&       O+&      56 \\
2&   19 27.4&  -14 32&     ...&              V4372 Sgr&    B2 IV&    38 \\
3&   16 12.8&  +12 24&    PN IC 4593&           ...&        ...&     25  \\
4&   15 51.0&  +15 05&    ...&             PG 1548+149&      DA&     18  \\
5&   19 54.9&  -13 48&   PN NGC 6818&           ...&        ...&     22 \\
6&  20 11.0&  -08 48&      ...&             HD 191639&      B1 V&    4  \\
7&   18 12.7&   +07 15& PN NGC 6572&            ...&        ...&     25 \\
8&    22 28.7&  -20 40& PN NGC 7293&           ...&         ...&    17 \\
9&    21 04.6&  -11 40& PN NGC 7009&             ...&       ...&     19 \\
10&   18 18.0&  +09 53& PN PHL 932&              ...&       ...&     16 \\
11&   21 09.6&  -10 07&  ...&                   ...&       ...&       ... \\
12&   16 44.4&  +23 44& PN NGC 6210&             ...&        ...&     5 \\
13&   21 58.1&  -11 55&     ...&                 ...&       ...&    ... \\
14&   17 41.7&  +21 41&  PN G045.6+24.3 (K1-14)& ...&       ...&     19 \\
15&   20 14.1&  +12 44&  PN NGC 6891&           ...&       ...&      16 \\
16&   20 26.4&  +11 54&     ...&                 ...&      ...&      ... \\
17&   16 29.3&  +35 15&     ...&                  ...&     ...&      ... \\
18&   20 11.8&  +13 27&   PN IC4997&              ...&     ...&       5 \\
19&   18 20.3&   +38 02&    ...&           FBS 1815+381&    DA&       35 \\
20&   18 49.9&   +39 03&    ...&                 ...&      ...&      ... \\
21&   21 12.9&  +20 44&     ...&                  ...&     ...&       ... \\
22&   21 50.4&  +12 57& PN LBN 069.67-30.36&    ...&        ...&      16 \\
23&   18 46.4&  +42 05&      ...&                ...&       ...&      ... \\
24&   21 16.5&  +24 20&  PN PK072-17 (A74)&      ...&       ...&       7 \\
25&   22 20.7&  +11 56&      ...&               31 Peg &   B2 IV-Ve&   21 \\
26&   13 47.0&  +39 38&       ...&              ...&        ...&       ... \\
27&   19 46.1&  +50 46&  PN NGC 6826&           ...&        ...&       19 \\
28&   19 24.9&  +60 40&     ...&                 ...&       ...&        ... \\
29&   18 01.1&  +66 34&  PN NGC 6543&            ...&       ...&       15 \\
30&   23 26.1&  +42 19&   PN NGC 7662&           ...&       ...&       13 \\
31&   20 25.3&  +76 40&       ...&          KUV20417+7604&   sd:O&       60 \\
32&   23 34.3&  +49 04&  PN PP105 (K1-20)&       ...&        ...&      51 \\
33&   13 54.6&  +77 33&       ...&                ...&       ...&      ... \\
34&   00 46.9&  -11 51&    PN NGC 246&            ...&       ...&        2 \\
35&   23 38.2&  +81 07&   PN LBN 120.29+18.39&  ...&          ...&      26 \\
36&   12 39.7&  +82 19&   PN IC 3568&            ...&         ...&       20 \\
37&   01 00.7&  +16 08&   PN PHL 932&              ...&       ...&       26 \\
38&   01 10.7&  +09 58&        ...&             PG 0108+101&   DO&       25  \\
39&   03 13.4&  +72 44&        ...&                ...&       ...&       ... \\
40&   01 24.9&  +21 28&        ...&             PG 0122+214&   Sd?&      12 \\
41&   01 57.0&  +36 25&        ...&               ...&       ...&        ... \\
42&   10 49.9&  +67 16&        ...&               ...&       ...&        ... \\
43&   11 55.6&  +54 08&        ...&             EGGR 435&     DA&       37 \\
44&   02 09.9&  +31 04&        ...&               ...&       ...&       ... \\
45&   04 53.3&  +68 30&  PN IRAS 04488+6903&       ...&       ...&      28 \\
46&   09 25.8&  +69 05&         ...&           PG 0931+691&   sd:O&     57 \\
47&   02 37.3&  +34 11&        ...&          KUV 02335+3343&  sdB&      18\tablenotemark{a} \\
48&   11 11.7&  +54 52&   PN NGC 3587&             ...&       ...&      29 \\
49&   06 19.6&  +55 31&   PN G158.9+17.8 (PW1)&    ...&       ...&       6 \\
50&   08 42.6&  +58 02&  PN G158.8+37.1 (A28)&     ...&       ...&       14 \\
51&   03 27.8&  +27 27&         ...&               ...&      ...&       ... \\
52&   03 22.7&   +25 06&        ...&               ...&     ...&          ... \\
53&   02 42.1&   +10 05&        ...&               ...&      ...&       ... \\
54&   09 48.4&   +53 04&         ...&               ...&     ...&        ... \\
55&   03 58.7&   +30 37&         ...&             X Per&     O9.5pe&      50 \\
56&   07 59.3&   +53 33&  PN G164.8+31.1&           ...&      ...&        15 \\
57&   03 59.6&   +32 44&    PN NGC 1514&            ...&      ...&        4 \\
58&   05 55.2&    +45 51&   PN IC 2149&              ...&    ...&        19 \\
59&   06 40.1&   +40 37&         ...&           WD 0632+40&   DA&         45 \\
60&   06 37.4&   +33 27&        ...&               ...&      ...&         ... \\
61&   09 02.6&   +33 01&          ...&           CBS 90&      DA&         35 \\
62&   04 48.6&   +04 36&   PN HS 0444+453&          ...&      ...&       35 \\
63&   04 16.9&   -03 34&          ...&               ...&     ...&       ... \\
64&   05 05.9&   +00 47&          ...&               ...&     ...&       ... \\
65&   07 29.3&   +13 32&  PN G205.1+14.2 (YM29)&     ...&     ...&        18 \\
66&   07 31.0&   +06 48&        ...&               ...&      ...&         ... \\
67&   05 25.9&   -12 39&     PN IC 418&               ...&     ...&       23 \\
68&   02 30.4&   -26 12&         ...&                 ...&     ...&       ... \\
69&   08 42.2&   +09 26&         ...&                ...&     ...&         ... \\
70&   08 53.9&   +09 19&   PN PK 219+31 (A31)&     ...&       ...&       25 \\
71&   03 31.5&   -25 47&   PN NGC 1360&             ...&      ...&       25 \\
72&   05 55.6&   -18 21&          ...&              ...&      ...&        ... \\
73&   10 15.5&   +06 17&        ...&            WD 1010+064&   DA&         31 \\
74&   09 16.7&   -03 49&        ...&            PG 0914+037&   sd:O&    10 \\
75&   09 54.6&   -05 04&         ...&                ...&     ...&       ... \\
76&   10 24.6&   -18 19&    PN NGC 3242&             ...&      ...&       20 \\
77&   11 18.4&   -04 50&         ...&               ...&       ...&       ... \\
78&   12 23.5&   -19 23&     PN NGC 4361&           ...&       ...&       22 \\
79&   12 52.3&   -22 59&   PN G303.6+40.0 (A35)&   ...&       ...&       18 \\
80&   13 05.8&   +04 26&         ...&                ...&      ...&         ... \\
81&   13 34.3&   -25 22&        ...&           EC 13331-2540&  sd:B&       40 \\
82&   14 22.6&   -09 08&        ...&            G 124-26&       DA&       17 \\
83&   15 57.9&   -26 07&         ...&             Pi Sco&     B1 V&        13 \\
84&   16 21.7&   -25 31&        ...&             Sig Sco&     B1 III&      8 \\
\enddata
\tablenotetext{a}{WD 0230+343 (DA) is within 52 arc minutes of beam 
center.}
\end{deluxetable}

\clearpage

\end{document}